\documentclass{article}
\usepackage{spconf,amsmath,graphicx}
\usepackage{amssymb}
\usepackage{booktabs}
\usepackage{algorithm}
\usepackage{bbold}
\usepackage{algpseudocode}
\usepackage{amsfonts}
\usepackage{bbm}

\title{Big model only for hard audios: Sample dependent Whisper model selection for efficient inferences}
\name{Hugo Malard$^1$, Salah Zaiem$^2$, Robin Algayres$^{1,3}$}
\address{$^1$ ENS-PSL, Paris  $^2$LTCI, Télécom Paris, Institut Polytechnique de Paris \\ $^3$Inria, Paris  \\}
\begin{document}
\ninept
\maketitle
\begin{abstract}
Recent progress in Automatic Speech Recognition (ASR) has been coupled with a substantial increase in the model sizes, which may now contain billions of parameters, leading to slow inferences even with adapted hardware. In this context, several ASR models exist in various sizes, with different inference costs leading to different performance levels. Based on the observation that smaller models perform optimally on large parts of testing corpora, we propose to train a decision module, that would allow, given an audio sample, to use the smallest sufficient model leading to a good transcription. We apply our approach to two Whisper models with different sizes. By keeping the decision process computationally efficient, we build a decision module that allows substantial computational savings with reduced performance drops.

\end{abstract}
\begin{keywords}
Speech recognition, efficiency
\end{keywords}
\section{Introduction}
\label{sec:intro}

\begin{figure}[htb]
\begin{minipage}[b]{.48\linewidth}
  \centering
  \centerline{\includegraphics[width=3.88cm]{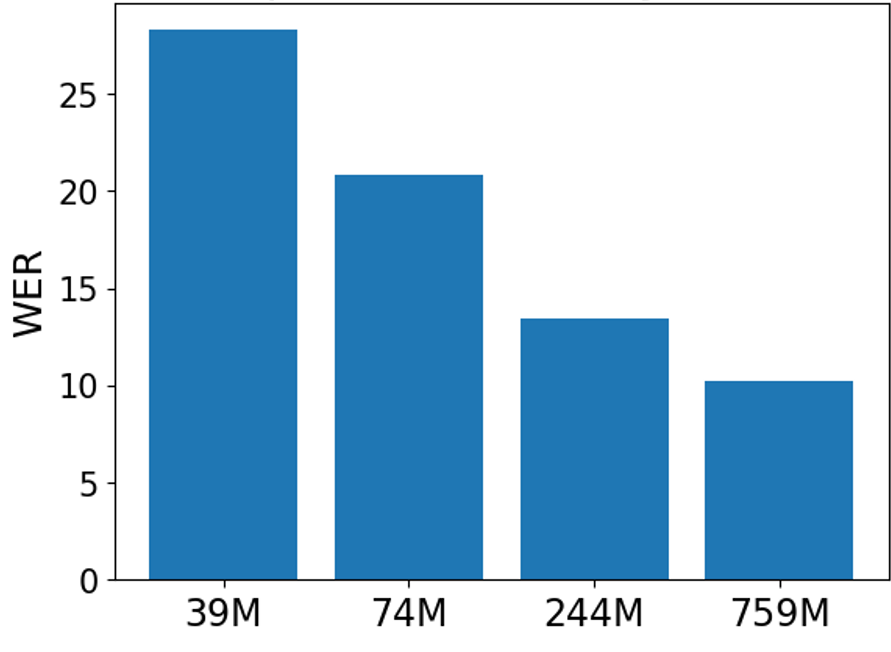}}
  \centerline{(a) Absolute performances}\medskip
\end{minipage}
\hfill
\begin{minipage}[b]{0.48\linewidth}
  \centering
  \centerline{\includegraphics[width=3.75cm]{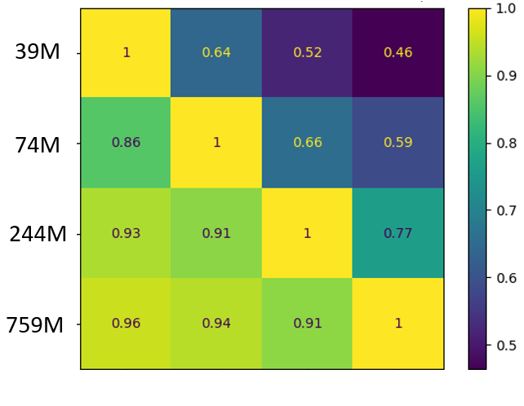}}
  \centerline{(b) Relative performances}\medskip
\end{minipage}
\vspace{-1.em}
\caption{Absolute and relative performances of Whisper models on CommonVoice test set. The four models are Whisper Tiny (39M parameters), Base (74M), Small (244M) and Medium (759M). Each cell $i,j$ in Figure (b) represents the percentage of utterances where the model $i$ performs at least as good as the model $j$. }
\label{fig:res}
\end{figure}

Recent progress in neural-based automatic speech recognition (ASR) has been driven by new modelling architectures, data collection and processing but also by larger models that have recently exceeded a billion parameters \cite{whisper}. Such advances have promised enhanced accuracy and capabilities, yet they have also ushered in escalating computational demands. These ASR models are usually available in a certain range of sizes with varying performance levels. For instance, Whisper models \cite{whisper} are available in 6 sizes from Tiny (39M parameters) to Large (1.5B parameters), Nvidia FastConformers \cite{rekesh2023fast} range from Large (118M parameters) to XXLarge (1.2B parameters) and self-supervised models like Hubert \cite{hsu2021hubert} or WavLM \cite{chen2021wavlm} are generally available in Base and Large versions. Systematically, following the deep learning trend across modalities, larger version models, although generally trained on the same datasets, perform substantially better than their reduced-size counterparts. This is shown in Figure \ref{fig:res} (a), where the mean Word Error Rates (WER) of four Whisper models with different sizes on the test set of CommonVoice \cite{CV} are presented. The mean WER drops from $28.1$ with the ``Tiny" version to $10.2$ with the ``Medium" one.

However, as shown in Figure \ref{fig:res} (b), this performance drop may not concern a significant part of the testing points. In this figure, every cell $(i,j)$ shows the proportion of samples in the CommonVoice test set where model $i$ performs better or equally to model $j$. For instance, the third cell in the first line (cell $(0,2)$) states that for $52\%$ of the testing samples, the Tiny model (39M) performs equally or better than the Small one (244M) while bearing more than $6$ times fewer parameters. Based on this observation, this work explores whether we can predict if audio samples will fall into this category. By doing so, audio samples that would not benefit from the costly inference of a large model can be assigned to a smaller one in order to reduce the total computational load. 

More precisely, this study aims to develop a \textit{decision module} that, given an audio sample, chooses a Whisper model version that has the lowest inference cost without WER degradation. Due to the complexity of the task, in this paper, we only focus on deciding if an audio sample should be decoded with Whisper Tiny (39M) or with Whisper Small (244M). These two model versions are relevant candidates as they exhibit high differences in both WER, inference costs and latency. 

A few works \cite{ewer2, ewer3} have already attempted to choose among several ASR model versions using WER prediction. Given the textual output of an ASR model, they explored the prediction of the sentence-level WER. Yet, these methods are not aiming to reduce inference costs but rather to decide whether an audio sample should be retreated by a more complex ASR model. Indeed, the most efficient techniques predict WER using full ASR pipelines based on either acoustic encoding and language model (LM) beam search \cite{ ewer2, ewer3}. Such methods that rely on costly beam searches cannot be used in our case where the aim is to reduce the computational load. 

Another close line of work is dynamic or early-exiting approaches. Instead of saving computation by choosing between separate ASR models, these methods have attempted to make forward passes lighter by skipping some of the last transformer layers of an ASR model \cite{salahrobin, yoon2022hubertee}. Decision to exits is based on entropy or representation-similarity thresholds. However, early-exiting, as developed in these works, can only save layer computation in the ASR encoder, while, as shown in Table \ref{tab:compute}, for attentional encoder-decoder architectures, most of the computations occur in the beam search decoding. 

The closest work to our effort is from Lugosch \textit{et al.} \cite{surprisal}, who propose to save computation cost at inference time by choosing between a large and a small ASR decoder. However, their method relies only on the log-likelihood of the encoder which is one of the baselines of our work. 

This paper explores possibilities to build a decision module that allows efficient selection between different ASR model sizes while keeping high performance. Our contributions are threefold : 
\begin{itemize}
    \item We successfully reduce inference costs for a negligible WER degradation. In addition, our method can be used to interpolate between model sizes, saving the need for costly training of intermediate models.
    \item We explore different inputs and architectures for the decision module and compare them to several baselines and toplines. 
    \item The codebase\footnote{https://github.com/hugomalard/Big-model-only-for-hard-audios.git}, developed within the SpeechBrain \cite{ravanelli2021speechbrain} framework,  is released for further investigations. 
\end{itemize}

\section{Methods}\label{seq:methods}

\begin{figure}[htb]
    \begin{center}
  \includegraphics[width=0.8\linewidth]{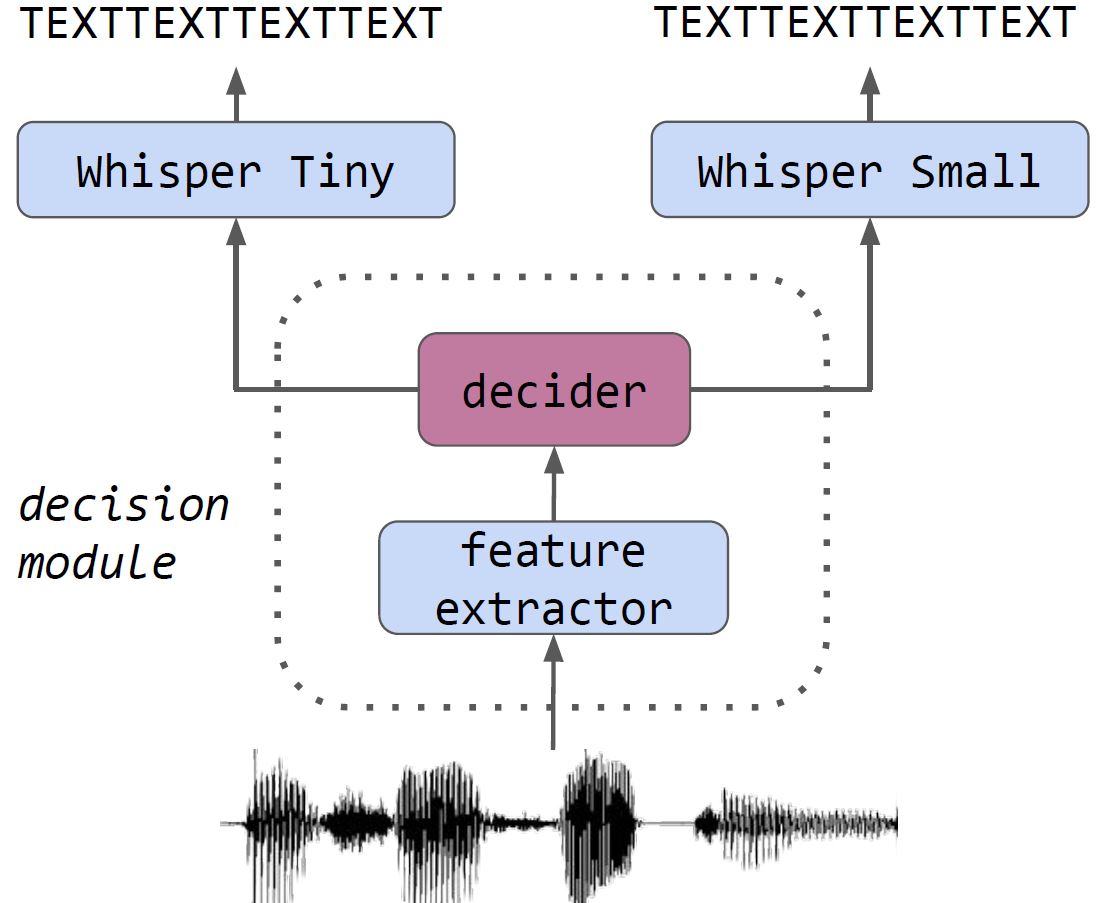}
  \vspace{-1.em}
  \caption{The decision module is composed of a feature extractor that encodes the speech signal into latents that are given in input to the decider. Based on the output of the decider, the audio sample is transcribed either by Whisper Tiny or Whisper Small.}
  \label{fig:deciderLatent}
  \end{center}
\end{figure}

This section describes our main pipeline represented in Figure \ref{fig:deciderLatent}. For a given audio sample, we first extract speech features and connect the output to a decider module. This latter is responsible for choosing to transcribe the audio with either a cheap model, Whisper Tiny, or a more expensive one, Whisper Small. 

\subsection{Feature extractor}
One of the questions we tackle in this work is which speech features are needed to predict how hard to transcribe is an audio sample. We explore two levels of representations, depending on their closeness to the raw signal waveform. We call low-level features hand-crafted representations like Mel spectrograms or MFCCs while high-level features are typically the outputs of a transformer encoder trained with self-supervision \cite{chen2021wavlm,W2V2} or text supervision \cite{whisper}. Intuitively, it is reasonable to think that high-level features will provide better input to the decider module. Indeed, while low-level features will only provide a frequency analysis of the speech input, higher-level features provide high-quality zero-shot phonetic encoding \cite{riviere2020unsupervised} and contain various additional information such as speaker and gender identity \cite{de_Seyssel_2022} as well as audio backgrounds \cite{whisperAT}. 

However, regarding our objective of minimizing computational cost, low-level features may be a favorable choice compared to high-level ones. Yet, we observed that the computational cost of representing speech with high-level features is negligible compared to the cost of the full ASR pipeline. Indeed, attention-based encoder-decoder models, like Whisper, are composed of an encoder and a decoder model that have virtually the same number of parameters. The encoder represents speech into latents and the decoder turns the latents into text with a beam search. While the encoding only costs one forward pass through the encoder, the decoding costs a large number of forward passes through the decoder. For instance, we present in Table \ref{tab:compute} the computational cost of encoding and decoding the test set of the CommonVoice \cite{CV} with either Whisper Tiny or Whisper Small. As expected, it appears that most of the computation is done in the beam-search decoding.

Therefore, considering both the quality of high-level features and the relatively low cost of encoding speech with a transformer stack, we decided to use the Whisper Small encoder as our feature extractor. We ablate this choice in the results section at Table \ref{AccuracyTab}.
\begin{table}[t]
\begin{center}
\resizebox{0.8\linewidth}{!}{
\begin{tabular}{lcc}
\toprule
\bf Model &\bf Encoding & \bf Beam-search decoding  \\
\toprule
Whisper Tiny & 0.01T & 0.22T\\
Whisper Small & 0.13T &  2.49T\\
\bottomrule
\end{tabular}
}
\end{center}
\vspace{-1.em}
\caption {Total MAC (multiply–accumulate) operations on CommonVoice test set for the encoder and the decoder of Whisper Tiny and Small using a beam search with n=8}
\label{tab:compute}
\vspace{-1.em}
\end{table}


%

\subsection{The decider}

As shown previously in Figure \ref{fig:res} (b), Whisper Tiny performs as well or even better than Whisper Small in $52\%$ of the sentences in the CommonVoice dataset. The decider from Figure \ref{fig:res} is a neural model that exploits this observation. Specifically, for an audio sample $a$, the decider is trained on the output of the frozen feature extractor to predict 1 or 0 according to the following equation.

\begin{equation}\label{eq:obj}
g_{\mathcal{M}_{T},\mathcal{M}_{S}}(a)=\mathbbm{1}_{WER(\mathcal{M}_{T}(a)) > WER(\mathcal{M}_{S}(a))}
\end{equation} 

Where $\mathcal{M}_{T}$ and $\mathcal{M}_{S}$ are respectively Whisper Tiny and Whisper Small. It means that the model should learn to predict 0 is the true WER obtained with  $\mathcal{M}_{T}$ is lower than the one of $\mathcal{M}_{S}$.

In a computationally aware context, using a lightweight module as decider is crucial, as its computational cost will be systematically paid at each inference. Using a transformer stack for the decider would induce a high inference cost due to the high dimensionality of the output of the feature extractor. A convolutional stack, scaling linearly with the sequence length, does not suffer from those side effects. Moreover, the locality bias of the convolutions may be pertinent in this context since errors may occur at localized segments of the speech sample. Therefore, we opted for a one-dimensional small ResNet \cite{He2015DeepRL} for the decider module architecture.

Instead of simply connecting the feature extractor output to the decider input, we learn a weighted sum of the feature extractor layers as in \cite{yang2021superb}. This method exploits the fact that different layers of a transformer stack encode different types of information from the audio signal \cite{zaiem2023speech}.

Here is a summary of our pipeline at inference time for transcribing one given audio sample, $a$. First, $a$ is turned into latents using Whisper Small encoder. The decider computes a learned weighted average of the encoder layers before doing a forward pass in a small 1D ResNet. If the decider output value falls below a threshold $h$, the beam search is performed in the Whisper Small decoder. Otherwise, $a$ is encoded and beam-search decoded with Whisper Tiny.




\section{Experiments}


In this section, we describe the datasets and hyperparameters used for the training of the decider model. In addition, we present baselines and toplines that act as comparison points to our method.
 %
\vspace{-1.em}
\subsection{Datasets and setups}
 
In this study, two datasets are considered: LibriSpeech \cite{libri} and CommonVoice 7.0 \cite{CV}. The LibriSpeech corpus is composed of read English speech recordings with 960 hours for training, two dev splits \textit{dev-clean} and \textit{dev-other} and two test splits \textit{test-clean} and \textit{test-other} of 5 hours each. CommonVoice is a collection of speech samples from worldwide users recording themselves with their own devices covering a large variety of age, gender and accents. The English part of the dataset presents roughly 1260 hours of recorded audio.

Our decider module is a small ResNet \cite{He2015DeepRL}, with 3 ResBlocks of two convolutional layers, each with 256 feature maps. The output of the ResNet is average-pooled before going through a linear layer with one sigmoid output neuron. In addition, the decider has one learnable weight per feature extractor layer. The feature extractor remains frozen during the decoder training. Training is done with binary cross entropy using Adam\cite{kingma2017adam} optimizer and a learning rate of $10^{-5}$ with cosine annealing.


\subsection{Baselines and Toplines}

In order to compare our decision module from Section \ref{seq:methods} to simpler methods with threshold-based decisions, we present in this subsection three considered baselines.

First, based on the well-known impact of noise on ASR quality, the output of a blind Signal to Noise Ratio (SNR) estimator \cite{brouhaha} is used for the decision module. Here, the decider relies on a simple threshold, determined by equal error rate: it runs Whisper Small if the computed SNR is lower than the threshold, otherwise it uses Whisper Tiny.
A second baseline consists of using an accent detection model \cite{zuluaga} to assign audios of rarer accents to the larger model. Indeed, \cite{Ngueajio_2022} shows a significant effect of English accents on ASR performances. For this baseline, the decision module consists in selecting Whisper Tiny if the English accent detected is either American, British or Canadian.

Finally, inspired by Lugosch \textit{and al.} \cite{surprisal}, a third baseline explores the use of ASR encoder-decoder logits as a confidence measure. Precisely, using Whisper Tiny, we perform a full greedy decoding, compute the entropy of the logit probabilities for each time-step, then aggregate  with mean pooling over the time dimension. Here again, the decider is a threshold on entropy values that is set on a validation set as for the SNR baseline.

Regarding toplines, we propose two decision modules that are voluntarily unrealistic in order to show the potential computational savings in an ideal case scenario. The first topline is an oracle that knows the true value of equation \ref{eq:obj} for any audio sample. For the second topline, we assume that the WER of Whisper Tiny is known in advance for any audio sample. The decider here is a threshold on WER values set on a validation set to determine when to use the larger Whisper model. The threshold is chosen such that it corresponds to the larger value for which there is no false negative (i.e choosing  ``Tiny" while it performs less well than ``Small").
\section{Results and Discussion}

\subsection{Accuracy of decision modules}

Table ~\ref{AccuracyTab} presents the decision modules accuracies on equation \ref{eq:obj}. It consists, for a collection of audio samples, in the percentage of times a decision module correctly assigns audio samples to the appropriate Whisper model as defined in Equation \ref{eq:obj}. 

The three first rows present the accuracy of the three baselines. The two first ones achieve accuracies close to random performance, which shows that neither SNR nor accent seems to capture the difficulty of the audio sample. For SNR, we hypothesize that this is due to the fact that the recordings come from relatively clean backgrounds, leading to very noisy and poorly informative estimation from the blind SNR estimator. On the contrary, logit-level entropy, although being computationally costly gives significantly better results reaching $64.5\%$ accuracy. It seems to indicate that the model internal states contain useful information about the difficulty of decoding audio samples. 

The second part of the table considers different feature extractors to our ResNet decider. As expected, higher-level features perform better than 80-dimensional Mel Spectrograms. The encoder of Whisper Small model performs better than the Wav2vec2.0 in our setting with $68.2\%$ accuracy, compared to $63.9\%$ for Wav2Vec2.0 features. This suggests that model-related features are the best adapted to the decision task. 

Finally, the oracle based on a threshold over the WER of Whisper Tiny produces strong results, showing that the WER of smaller Whisper version can be used to allocate audio samples to large models efficiently. However, blind WER prediction remains a noisy and computationally costly endeavour \cite{ewer3}.

We ablate the architectural choice of the decider model in Table \ref{tab:ablation}. First, removing the learned weighted sum of encoder layers slightly degrades the accuracy. Second, using a one-layer transformer network with a roughly equal number of parameters to the ResNet, apart from raising the cost of the decision, also degrades accuracy scores by a couple of percent. Finally, inspired by \cite{whisperAT}, we implemented a small TL-transformer architecture which is composed of one transformer layer on the encoder layers outputs and another one on their pooled (time-wise) representations. This approach scores even lower than the simple one-layer transformer architecture. 

\subsection{WER/MACs trade-off}

Table \ref{tab:macs} shows the trade-off WER/MACs obtained with our pipeline compared to transcribing speech using Whisper models. MACs, which stands for multiply–accumulate operations, is our measure of computational cost. It is important to note that the MACs column shows the cost of the full pipeline for transcribing the CommonVoice test set, including  the cost of the decision module when there is one. 
Table \ref{tab:macs} starts with the WER/MACs of simply transcribing speech using the different Whisper models with beam search. Then, we include the logit entropy of Whisper Tiny, which is the best baseline from Table \ref{AccuracyTab}. This baseline increases the WER by an absolute $0.7$ points while reducing the model computational cost by $150G$. Comes next, our main pipeline that uses a decision module composed of Whisper Small encoder and Resnet. For this latter, we provide scores at two different thresholds, $0.3$ and $0.5$. By comparison with simply running Whisper Small, selecting a threshold of $0.5$ on the sigmoid output  of our decision module gives a $16\%$ higher WER while resulting in a $35\%$ decrease in MACs. Using a threshold of $0.3$ increases the WER by an absolute $0.37$ points while reducing the model computational cost by $310G$ MACs (\textit{i.e.} $12\%$ of the total cost). 
Finally, the last 2 lines show the hypothetical improvements that can be obtained using a perfect decider, or a threshold, based on a perfect estimation of the WER of $\mathcal{M}_T$. Not only they allow to reduce the MACs, but they also reduce significantly the computational load. These toplines confirm the potential of the approach and call for further research on model size assignment.

\begin{figure}[h]
\begin{minipage}[b]{1.0\linewidth}
  \centering
  \centerline{\includegraphics[width=6cm]{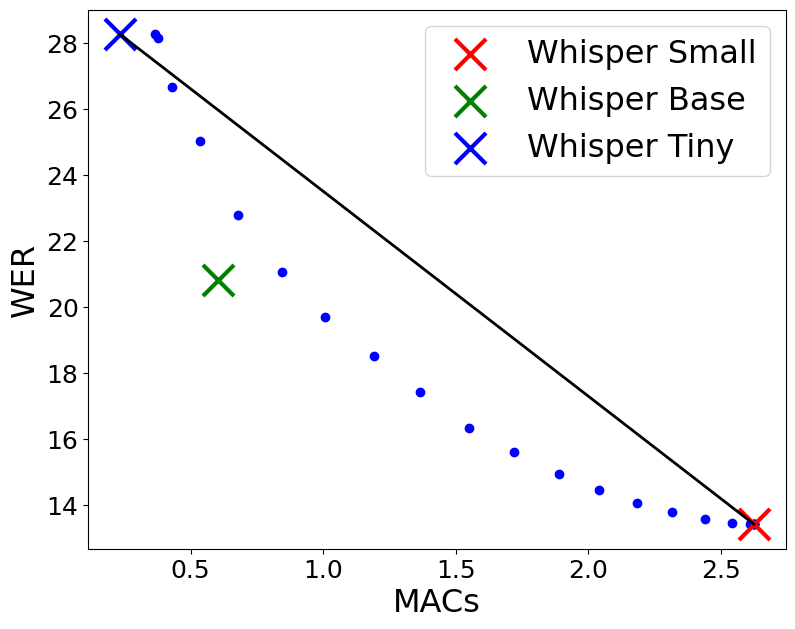}}
\end{minipage}
\caption{WER/MACs values on the CommonVoice test set for different pipelines. The blue dots are the values for multiple thresholds using our best decision module (Whisper Small encoder and ResNet). The crosses correspond to the Whisper models Tiny, Base and Small respectively. Finally, the black line is a linear interpolation between the MACs and WER of Whisper Tiny and Whisper Small}
\label{fig:macWer}
\end{figure}

Figure~\ref{fig:macWer} shows the performances (WER) and computational costs (MACs) of the intermediate models obtained using our main pipeline (i.e. Whisper Small encoder and ResNet). Almost all the points are  under the plotted diagonal, which means that the resulting drop in performance (relative to the larger model) is systematically smaller than the gain in computational cost. 
Whisper Base is included in Figure~\ref{fig:macWer} as it is an intermediate model between Whisper Tiny and Small. Its WER/MACs is only slightly better than our selection approach. This shows, that the method presented yields nearly equivalent performance to that of an intermediate model trained entirely from scratch, saving very costly training. 

    \begin{table}[t]
\begin{center}
\resizebox{\linewidth}{!}{
\begin{tabular}{llccc}
\toprule
\bf Feature extractor & \bf Decider &\bf test-clean$\uparrow$  & \bf test-other$\uparrow$  &\bf CV test$\uparrow$  \\
\toprule
SNR \cite{brouhaha}& thresh. & 50.7 & 47.2 & 47.0\\
Accent \cite{zuluaga} & thresh. & n/a & n/a& 52.0\\
$\mathcal{M}_{T}$ logit entropy & thresh. & 64.4 & 63.7 & 64.5\\
\midrule
Mel f-banks & ResNet &62.5 & 55.2 & 60.0 \\ 
Wav2Vec2.0 Base & ResNet &52.3 & 57.7 &63.9\\
$\mathcal{M}_{T}$ encoder& ResNet  & 65.1& 65.0 & 66.4 \\
$\mathcal{M}_{S}$ encoder& ResNet &\bf 68.0 & \bf 66.6 & \bf 68.2\\
\midrule
\it WER $\mathcal{M}_{T}$ &  \it thresh. & \it 84.8 &  \it 80.7 &  \it 80.3\\
\bottomrule
\end{tabular}
}
\end{center}
\caption {Accuracies (higher the better) of the several decision modules on LibriSpeech and CommonVoice test sets. SNR, Accent and Logit Entropy are baseline models that only require to fit a threshold on a validation set. When the Decider is ResNet, a dedicated ResNet is trained on each of the different feature extractors. The last line is a topline model based on an oracle that provides the WER of Whisper Tiny. $\mathcal{M}_{T}$ and $\mathcal{M}_{S}$ are respectively Whisper Tiny and Small.}
\label{AccuracyTab}
\end{table}

\begin{table}[h]
\begin{center}
\resizebox{0.8\linewidth}{!}{
\begin{tabular}{lcc}
\toprule
\bf Decider &\bf Test-Clean$\uparrow$ & \bf Test-Other$\uparrow$ \\
\toprule
ResNet  & \textbf{68.0} & \textbf{66.6} \\
\midrule
ResNet w/o weighted  & 66.4 & 66.0\\
Transformer  & 66.2 & 65.4 \\
TL-Transformer & 64.4 & 55.2\\
\bottomrule
\end{tabular}
}
\caption {Accuracy (higher the better) obtained by various decider models on LibriSpeech test sets using Whisper Small encoder as a feature extractor. 'w/o weighted' means that the decider does not learn a weighted average of the encoder layers. Transformer has one transformer layer and TL-transformer is a 1 layer replication of \cite{whisperAT}.}

\end{center}
\label{tab:ablation}
\end{table}

\begin{table}[h]
\begin{center}
\resizebox{0.85\linewidth}{!}{
\begin{tabular}{lcc}
\toprule
\bf Method & \bf WER$\downarrow$  & \bf MACs$\downarrow$ \\
\toprule
$\mathcal{M}_{T}$ & 28.1 & 0.23T \\ 
$\mathcal{M}_{Base}$& 20.7 & 0.60T\\
$\mathcal{M}_{S}$ & 13.3 & 2.62T\\
\midrule 
$\mathcal{M}_{T}$ logit entropy & 14.0 &2.47T\\
encoder $\mathcal{M}_{S}$ + ResNet @$0.5$ & 15.4 & 1.72T\\
encoder $\mathcal{M}_{S}$ + ResNet @$0.3$ & 13.7 & 2.31T\\
\midrule
\it WER $\mathcal{M}_{T}$ Oracle & \it 12.93 & \it 1.954T\\
\it Oracle & \it 12.27 & \it 1.468T\\
\bottomrule
\end{tabular}
}
\end{center}
\vspace{-1.em}
\caption {Average WER and MACs (lower the better) associated with each method on the CommonVoice test set. The table starts with the performances of $\mathcal{M}_{T}$, $\mathcal{M}_{Base}$ and $\mathcal{M}_{S}$ which are full ASR pipeline of respectively Whisper Tiny, Base and Small. $\mathcal{M}_{T}$ logit entropy is our baseline. encoder $\mathcal{M}_{S}$+ ResNet is our main contribution for which we give performances at two different theshold value (0.3 and 0.5). The last 2 lines are our toplines.}
\label{tab:macs}
\end{table}
\subsection{Discussion}
The failure of the baselines and the encoder layers being the best performing input, tend to show that the errors are very dependent of the ASR model, rather than complexities inherent to the audio signal that would make any ASR model fail to transcribe.

To investigate this, we compute correlation values between the WER of a Conformer Large \cite{conformer} model and of a Wav2Vec2.0 Base model fine tuned on LibriSpeech, with the WER of Whisper tiny, on the LibriSpeech test sets (combined clean and other).\\
The Pearson correlation coefficient between the WER of the Conformer model and the WER of Whisper Tiny reaches only $0.44$, while the Spearman correlation is only $0.41$. Similarly, these two correlation quantities between the WER of Wav2Vec2.0 and Whisper Tiny reach respectively $0.51$ and $0.45$. 
The low correlation values and the weak monotonic relationship seem to indicate that models have different intrinsic failure cases. It confirms, as our results have first shown, that a successful model selection approach needs model-related inputs.

\section{Conclusion}
In this study, we explored a new, computationally efficient approach, that selects for an audio sample the most efficient model among two of different sizes. It can be applied to interpolate between two trained models of fixed sizes without additional training, reducing the relative computational cost more than it degrades performances.

\newpage
\vfill\pagebreak

\section{Acknowledgements}
This work was funded in part by the European Research Council (ERC-2011-AdG-295810 BOOTPHON), the Agence Nationale pour la Recherche (ANR-17-EURE-0017 Frontcog, ANR-10-IDEX-0001-02 PSL*, ANR-19-P3IA-0001 PRAIRIE 3IA Institute) and grants from CIFAR (Learning in Machines and Brains), Meta AI Research (Research Grant), Google (Faculty Research Award), Microsoft Research (Azure Credits and Grant), and Amazon Web Service (AWS Research Credits). Furthermore, this work was performed using HPC resources from GENCI-IDRIS (Grant 2021-[AD011011217])

\bibliographystyle{IEEEbib}
\bibliography{strings,refs}

\begin{thebibliography}{10}

\bibitem{whisper}
Alec Radford, Jong~Wook Kim, Tao Xu, Greg Brockman, Christine McLeavey, and
  Ilya Sutskever,
\newblock ``Robust speech recognition via large-scale weak supervision,'' 2022.

\bibitem{rekesh2023fast}
Dima Rekesh, Nithin~Rao Koluguri, Samuel Kriman, Somshubra Majumdar, Vahid
  Noroozi, He~Huang, Oleksii Hrinchuk, Krishna Puvvada, Ankur Kumar, Jagadeesh
  Balam, and Boris Ginsburg,
\newblock ``Fast conformer with linearly scalable attention for efficient
  speech recognition,'' 2023.

\bibitem{hsu2021hubert}
Wei{-}Ning Hsu, Benjamin Bolte, Yao{-}Hung~Hubert Tsai, Kushal Lakhotia, Ruslan
  Salakhutdinov, and Abdelrahman Mohamed,
\newblock ``Hubert: Self-supervised speech representation learning by masked
  prediction of hidden units,''
\newblock {\em CoRR}, vol. abs/2106.07447, 2021.

\bibitem{chen2021wavlm}
Sanyuan Chen, Chengyi Wang, Zhengyang Chen, Yu~Wu, Shujie Liu, Zhuo Chen, Jinyu
  Li, Naoyuki Kanda, Takuya Yoshioka, Xiong Xiao, Jian Wu, Long Zhou, Shuo Ren,
  Yanmin Qian, Yao Qian, Jian Wu, Michael Zeng, and Furu Wei,
\newblock ``Wavlm: Large-scale self-supervised pre-training for full stack
  speech processing,''
\newblock {\em CoRR}, vol. abs/2110.13900, 2021.

\bibitem{CV}
Rosana Ardila, Megan Branson, Kelly Davis, Michael Henretty, Michael Kohler,
  Josh Meyer, Reuben Morais, Lindsay Saunders, Francis~M. Tyers, and Gregor
  Weber,
\newblock ``Common voice: A massively-multilingual speech corpus,'' 2020.

\bibitem{ewer2}
Ahmed Ali and Steve Renals,
\newblock ``Word error rate estimation without asr output: e-wer2,'' 2020.

\bibitem{ewer3}
Shammur~Absar Chowdhury and Ahmed Ali,
\newblock ``Multilingual word error rate estimation: e-wer3,'' 2023.

\bibitem{salahrobin}
Salah Zaiem, Robin Algayres, Titouan Parcollet, Slim Essid, and Mirco
  Ravanelli,
\newblock ``Fine-tuning strategies for faster inference using speech
  self-supervised models: A comparative study,'' 2023.

\bibitem{yoon2022hubertee}
Ji~Won Yoon, Beom~Jun Woo, and Nam~Soo Kim,
\newblock ``Hubert-ee: Early exiting hubert for efficient speech recognition,''
  2022.

\bibitem{surprisal}
Loren Lugosch, Derek Nowrouzezahrai, and Brett~H. Meyer,
\newblock ``Surprisal-triggered conditional computation with neural networks,''
  2020.

\bibitem{ravanelli2021speechbrain}
Mirco Ravanelli, Titouan Parcollet, Peter Plantinga, Aku Rouhe, Samuele
  Cornell, Loren Lugosch, Cem Subakan, Nauman Dawalatabad, Abdelwahab Heba,
  Jianyuan Zhong, et~al.,
\newblock ``Speechbrain: A general-purpose speech toolkit,''
\newblock {\em arXiv preprint arXiv:2106.04624}, 2021.

\bibitem{W2V2}
Alexei Baevski, Henry Zhou, Abdelrahman Mohamed, and Michael Auli,
\newblock ``wav2vec 2.0: A framework for self-supervised learning of speech
  representations,'' 2020.

\bibitem{riviere2020unsupervised}
Morgane Rivière, Armand Joulin, Pierre-Emmanuel Mazaré, and Emmanuel Dupoux,
\newblock ``Unsupervised pretraining transfers well across languages,'' 2020.

\bibitem{de_Seyssel_2022}
Maureen de~Seyssel, Marvin Lavechin, Yossi Adi, Emmanuel Dupoux, and Guillaume
  Wisniewski,
\newblock ``Probing phoneme, language and speaker information in unsupervised
  speech representations,''
\newblock in {\em Interspeech 2022}. sep 2022, {ISCA}.

\bibitem{whisperAT}
Yuan Gong, Sameer Khurana, Leonid Karlinsky, and James Glass,
\newblock ``Whisper-at: Noise-robust automatic speech recognizers are also
  strong general audio event taggers,'' 2023.

\bibitem{He2015DeepRL}
Kaiming He, X.~Zhang, Shaoqing Ren, and Jian Sun,
\newblock ``Deep residual learning for image recognition,''
\newblock {\em 2016 IEEE Conference on Computer Vision and Pattern Recognition
  (CVPR)}, pp. 770--778, 2015.

\bibitem{yang2021superb}
Shu{-}Wen Yang, Po{-}Han Chi, Yung{-}Sung Chuang, Cheng{-}I~Jeff Lai, Kushal
  Lakhotia, Yist~Y. Lin, Andy~T. Liu, Jiatong Shi, Xuankai Chang, Guan{-}Ting
  Lin, Tzu{-}Hsien Huang, Wei{-}Cheng Tseng, Ko{-}tik Lee, Da{-}Rong Liu, Zili
  Huang, Shuyan Dong, Shang{-}Wen Li, Shinji Watanabe, Abdelrahman Mohamed, and
  Hung{-}yi Lee,
\newblock ``{SUPERB:} speech processing universal performance benchmark,''
\newblock {\em CoRR}, vol. abs/2105.01051, 2021.

\bibitem{zaiem2023speech}
Salah Zaiem, Youcef Kemiche, Titouan Parcollet, Slim Essid, and Mirco
  Ravanelli,
\newblock ``Speech self-supervised representations benchmarking: a case for
  larger probing heads,''
\newblock {\em arXiv preprint arXiv:2308.14456}, 2023.

\bibitem{libri}
Vassil Panayotov, Guoguo Chen, Daniel Povey, and Sanjeev Khudanpur,
\newblock ``Librispeech: An asr corpus based on public domain audio books,''
\newblock in {\em 2015 IEEE International Conference on Acoustics, Speech and
  Signal Processing (ICASSP)}, 2015, pp. 5206--5210.

\bibitem{kingma2017adam}
Diederik~P. Kingma and Jimmy Ba,
\newblock ``Adam: A method for stochastic optimization,'' 2017.

\bibitem{brouhaha}
Marvin Lavechin, Marianne Métais, Hadrien Titeux, Alodie Boissonnet, Jade
  Copet, Morgane Rivière, Elika Bergelson, Alejandrina Cristia, Emmanuel
  Dupoux, and Hervé Bredin,
\newblock ``Brouhaha: multi-task training for voice activity detection,
  speech-to-noise ratio, and c50 room acoustics estimation,'' 2023.

\bibitem{zuluaga}
Juan Zuluaga-Gomez, Sara Ahmed, Danielius Visockas, and Cem Subakan,
\newblock ``Commonaccent: Exploring large acoustic pretrained models for accent
  classification based on common voice,'' 2023.

\bibitem{Ngueajio_2022}
Mikel~K. Ngueajio and Gloria Washington,
\newblock ``Hey {ASR} system! why aren't you more inclusive?,''
\newblock in {\em Lecture Notes in Computer Science}, pp. 421--440. Springer
  Nature Switzerland, 2022.

\bibitem{conformer}
Anmol Gulati, James Qin, Chung-Cheng Chiu, Niki Parmar, Yu~Zhang, Jiahui Yu,
  Wei Han, Shibo Wang, Zhengdong Zhang, Yonghui Wu, and Ruoming Pang,
\newblock ``Conformer: Convolution-augmented transformer for speech
  recognition,'' 2020.

\end{thebibliography}

\end{document}